\documentclass[aps,pra,twocolumn,showpacs,amssymb,amsmath]{revtex4}
\usepackage{graphicx}
\usepackage{dcolumn}
\usepackage{bm}

\begin{document}

\title{Geometric phase of a two-level system in a dissipative environment}

\author{Kazuo Fujikawa$^{1,2}$}
\author{Ming-Guang Hu$^1$}
\affiliation{$^1$Theoretical Physics Division, Chern Institute of
Mathematics, Nankai University, Tianjin 300071, People's Republic of
China\\$^2$Institute of Quantum Science, College of Science and
Technology, Nihon University, Chiyoda-ku, Tokyo 101-8308, Japan}
\date{\today}
\begin{abstract}
The geometric (Berry) phase of a two-level system in a dissipative
environment is analyzed by using the second-quantized formulation,
which provides a unified and gauge-invariant treatment of adiabatic
and nonadiabatic phases and is thus applicable to a quantitative
analysis of transitional regions away from ideal adiabaticity. In
view of the recent experimental observation of the Berry phase in a
superconducting qubit, we illustrate our formulation for a concrete
adiabatic case in the Ohmic dissipation. The correction to the total
phase together with the geometry-dependent dephasing time is given
in a transparent way. The behavior of the geometric phase away from
ideal adiabaticity is also analyzed in some detail.\end{abstract}


\maketitle

\section{Introduction}
A superconducting nanostructure \cite{2008Clarke} with its potential
scalability leads to a promising solid-state platform for quantum
information processing
\cite{
2007Majer,2007sillanpaa}. The coherent control of macroscopic
quantum states in superconducting circuits
\cite{2001Makhlin,2002Vion}, in particular two-level quantum
systems, makes it possible to observe the geometric phases (GPs). If
the evolution is adiabatic, GP is known as the Berry phase (BP)
\cite{1984Berry}, which arises from the cyclic evolution of a
quantum mechanical system and depends only on the area spanned in
the parameter space \cite{1992Anandan}. The experiments to observe
BP have been implemented recently by using the superconducting qubit
\cite{2007Leek,2008Mottonen}. It is intriguing that a
geometry-dependent dephasing was detected \cite{2007Leek} and it in
general indicates a coupling of the qubit with its environment,
which plays a non-negligible role in the superconducting circuits
\cite{1999Nakamura}. Given the argument that BP has an intrinsic
fault-tolerant robustness and is potentially used as the geometric
quantum logic gate \cite{1999Zanardi}, the environmental effects on
BP need to be quantitatively investigated. This issue of intrinsic
robustness is very important, but it appears that no consensus on
this issue yet \cite{2003Blais}.

The effects of environment on GPs have been analyzed by several
authors from various points of view: for example, the classic field
fluctuation \cite{2003Chiara}, the quantum jump \cite{2003Carollo},
GP distributions \cite{2004Marzlin}, etc.. For macroscopic quantum
states of superconducting circuits, the behavior of environment is
generally simulated by an infinite number of harmonic oscillators  with some definite spectrum
distribution of frequency \cite{1981Caldeira}. Specifically, for a
single qubit interacting with its environment, it is effectively
described by a spin-boson model, which has been used in the past to
analyze the quantum  decoherence due to dissipation
\cite{1981Caldeira,1987Leggett}. This model can also be used for the
analysis of GPs in dissipative environment and this direction has
been initiated in \cite{2003Whitney,2005Whitney}.

In this paper we exploit a general method for treatment of the
environmental effects in the interaction picture as a correction to
effective energy eigenvalues appearing in the evolution operator. We
here recall that the geometric phases, either adiabatic or
nonadiabatic, are associated with the time development of the state
vector typically during one cyclic evolution. The time development
of the state vector is described by the Schr\"{o}dinger equation,
and the time development is entirely generated by the Hamiltonian.
The evolution operator, which is generically defined by
$|\psi(t)\rangle=U(t,0)|\psi(0)\rangle$, if evaluated exactly thus
contains all the information about the geometric phases. The primary
object of our study is thus the evolution operator.

To the second-order perturbation, we give a formula for studying the
effects of the environment on GPs for a spin in the magnetic field
at zero temperature, $T=0$. Compared with the previous relevant work
\cite{2003Whitney}, our formulation is valid for adiabatic
\cite{1984Berry} and nonadiabatic (cyclic or non-cyclic)
\cite{1987Aharonov,1988Samuel} evolutions. It shows the effects from
a different point of view and is easily extended to analyze the
dissipation-related issues such as adiabatic quantum computation
\cite{2001Farhi} and quantum decoherence \cite{2006Nombardo}. In the
following we first derive the general formula and then take the
adiabatic limit to analyze the environmental effects on BP in a
concrete manner. The modification of GPs by environmental effects in
the regions away from ideal adiabaticity is also analyzed.

This article is organized as follows. In Sec. \ref{sec-review}, we
review the second-quantized formulation of GPs in the absence of
environment, which is exact for a single-spin system in magnetic
field. Based on the formulation, we take the effects of environment
into consideration in Sec. \ref{sec-GP}, in which a perturbation
correction to GPs originating from the coupling with environment is
displayed. As illustrations, in Secs. \ref{subsec-Adiabatic} and
\ref{subsec-nonadiabatic} we show the effects of environment on
adiabatic Berry phase and nonadiabatic geometric phase,
respectively. Finally, Sec. \ref{sec-con} is dedicated to
conclusion.

\section{Review of the second-quantized formulation of geometric
phases \label{sec-review}} In this section, let us review the basic
idea of the second-quantized formulation of GPs without taking
dissipation into account. It was introduced by one of the present
authors in
\cite{2005Deguchi,2005Fujikawa,2007Fujikawa,2008Fujikawa}. Instead
of describing a general theory, we consider the GPs of a spin in
rotating magnetic field to demonstrate how the method works. Note
that such a situation was already discussed in
\cite{2007Fujikawa,2008Fujikawa}.

We denote a rotating background (magnetic) field by ${\bm
B}(t)=B\big(\sin\theta \cos\varphi(t),
\sin\theta\sin\varphi(t),\cos\theta\big)$ and $\varphi(t) =\omega_0
t$ with a constant angular velocity $\omega_0$. The action for the
system in the second-quantized formulation ($\hbar=1$) is
\begin{eqnarray}\label{eq-action1}
S&=&\int dt\left[\hat{\psi}^\dag(t)\Big(i\frac{\partial}{\partial
t}+{\bm B}\cdot {\bm \sigma}/2\Big)\hat{\psi}(t)\right],
\end{eqnarray}
where the field operator is expanded as
$\hat{\psi}(t,\vec{x})=\sum_{l=\pm}\hat{c}_l(t)w_l(t)$ with the
anti-commutation relation,
$\big\{\hat{c}_l(t),\hat{c}_m^\dag(t)\big\}=\delta_{lm}$. For later
use, here we define the Fock states by
$|l\rangle=\hat{c}^\dag_l(0)|0\rangle$ with the vacuum state
$|0\rangle$ satisfying $\hat{c}_l(0)|0\rangle=0$. We use only the very
elementary aspect of the second quantization of the fermion to
clarify the hidden gauge symmetry that controls all GPs.

For the above specific magnetic field with time-independent
$\theta$, the effective Hamiltonian for the {\em isolated} spin system is exactly diagonalized and, consequently, the isolated spin system is exactly solvable if
one chooses the basis vectors as
\begin{eqnarray}
w_{+}(t)=\left(\begin{array}{c}
             e^{-i\varphi(t)}\cos\frac{\vartheta}{2}\\
            \sin\frac{\vartheta}{2}
            \end{array}\right),\
w_{-}(t)=\left(\begin{array}{c}
             e^{-i\varphi(t)}\sin\frac{\vartheta}{2}\\
            -\cos\frac{\vartheta}{2}
            \end{array}\right),
\end{eqnarray}
with $\vartheta=\theta-\theta_0$ and the constant parameter
$\theta_0$ defined by
\begin{equation}\label{eq-xi}
\tan\theta_0=\frac{\omega_0 \sin\theta}{B+\omega_0\cos\theta}.
\end{equation}
Then we have
\begin{eqnarray}\label{eq-weigen}
&&w_{\pm}^{\dagger}(t)\hat{h}w_{\pm}(t)
=\mp \frac{1}{2}B\cos\theta_0,\nonumber\\
&&w_{\pm}^{\dagger}(t)i\partial_{t}w_{\pm}(t)
=\frac{1}{2}\omega_0[1\pm\cos(\theta-\theta_0)],
\end{eqnarray}
with $\hat{h}=-{\bm B}(t)\cdot {\bm \sigma}/2$. In the operator
formulation of the second-quantized theory, we obtain a diagonalized
effective Hamiltonian, $\hat{H}_{\rm{eff}}(t)=\sum_{l=\pm} E_l
\hat{c}_l^\dag(t) \hat{c}_l(t)$, where two time-independent
effective energy eigenvalues are given by
\begin{eqnarray}\label{eq-Epm}
E_\pm&=&w_{\pm}^{\dagger}(t')\big(\hat{h}
-i\partial_{t'}\big)w_{\pm}(t')\nonumber\\
&=&
\mp\frac{1}{2}B\cos\theta_0-\frac{1}{2}\omega_0\big[1\pm\cos(\theta-\theta_0)\big].
\end{eqnarray}
By noting the Heisenberg equation of motion
\begin{equation*}
i\frac{\partial}{\partial
t}\hat{c}_l(t)=[\hat{c}_l(t),\hat{H}_{\text{eff}}(t)],
\end{equation*}
it is confirmed that one can write
\begin{equation*}
\hat{c}_l(t)=U^{\dagger}(t)\hat{c}_l(0)U(t)
\end{equation*}
by introducing the ``Schr\"{o}dinger picture" effective Hamiltonian $\hat{\mathcal{H}}_{\text{eff}}(t)\equiv\sum_lE_l\hat{c}_l^\dag(0)\hat{c}_l(0)$ and the second-quantized formula of the evolution operator defined by
\begin{equation}\label{eq-evolution}
U(t)=\mathcal{T}\exp[-i\int^t_0\hat{\mathcal{H}}_{\text{eff}}(t')
dt']
\end{equation}
where $\mathcal{T}$ represents a time-ordering product. In general,
we have $\hat{\mathcal{H}}_{\text{eff}}(t)=
\sum_{l,m}E_{l,m}(t)\hat{c}_l^\dag(0)\hat{c}_m(0)$ for the
time-dependent $\hat{h}(t)$, and the adiabatic approximation
corresponds to an approximate diagonalization of
$\hat{\mathcal{H}}_{\text{eff}}(t)$. See
\cite{2005Deguchi,2005Fujikawa,2007Fujikawa, 2008Fujikawa} for more
details.

For the Schr\"{o}dinger equation
$i\partial_{t}\psi_{\pm}(t)=\hat{h}\psi_{\pm}(t)$ with initial
condition $\psi_\pm(0)=w_\pm(0)$, its {\em exact} solution is given
in the second-quantized notation \cite{2007Fujikawa, 2008Fujikawa}
\begin{eqnarray}\label{eq-exactamplitude}
\psi_{\pm}(t)&=&\langle
0|\hat{\psi}(t)\hat{c}^\dag_\pm(0)|0\rangle\nonumber\\
&=&\sum_lw_l(t)\langle 0|\hat{c}_l(0)U(t)\hat{c}^\dag_\pm(0)|0\rangle\nonumber\\
&=&w_{\pm}(t)\exp\left[-i\int_{0}^{t}dt'
w_{\pm}^{\dagger}(t')\big(\hat{h}
-i\partial_{t'}\big)w_{\pm}(t')\right],\nonumber\\
\end{eqnarray}
where the exponent has been calculated in Eq. (\ref{eq-Epm}). Since
$w_{\pm}(T)=w_{\pm}(0)$ with the period $T=2\pi/|\omega_0|$, the
solution is cyclic \cite{1987Aharonov} and, as an exact solution, it
is applicable to the nonadiabatic case also. For an arbitrary
time-dependent ${\bm B}(t)$, any exact solution of the
Schr\"{o}dinger equation can be written in the last form of
Eq.(\ref{eq-exactamplitude}) , if one chooses basis vectors
$w_{\pm}(t)$ suitably \cite{2007Fujikawa}. But the periodicity
$w_{\pm}(T)=w_{\pm}(0)$ is generally lost and thus the solution
becomes non-cyclic \cite{1988Samuel}.

Actually, at the adiabatic limit $|\omega_0/ B|\ll 1$, $\theta_0$ in
Eq. (\ref{eq-xi}) approaches zero so that the conventional BP $\pi
(1\pm\cos\theta)$ \cite{1984Berry} is recovered from Eqs.
(\ref{eq-weigen}) and (\ref{eq-exactamplitude}). On the other hand,
at the nonadiabatic limit $|\omega_0/ B|\gg 1$, $\theta_0$
approaches $\theta$ so that GP in Eq. (\ref{eq-exactamplitude})
vanishes. Namely, the adiabatic BP is smoothly connected to the
trivial phase inside the exact solution \cite{2008Fujikawa}. We can
thus analyze a transitional region from the adiabatic limit to the
nonadiabatic region, which was not possible in the past formulation.

One can assign a gauge-invariant meaning to the GP under general
adiabatic or nonadiabatic evolution. To see this, let us recall that
the field variable
$\hat{\psi}(t,\vec{x})=\sum_{l=\pm}\hat{c}_l(t)w_l(t)$ in
Eq.(\ref{eq-action1}) is invariant under the simultaneous
replacements \cite{2005Fujikawa}
\begin{equation}\label{eq-hiddenlocal}
\hat{c}_l(t)\rightarrow e^{-i\alpha_l(t)}\hat{c}_l(t),\ \
w_l(t)\rightarrow e^{i\alpha_l(t)}w_l(t),
\end{equation}
and thus basic action (\ref{eq-action1}) [even with dissipation; see
Eq.(\ref{eq-action})] is invariant under this exact gauge symmetry.
One then confirms that the exact Schr\"{o}dinger amplitude
$\psi_l(t)=\langle 0|\hat{\psi}(t)\hat{c}^{\dagger}_l(0) |0\rangle$
in Eq.(\ref{eq-exactamplitude}) is transformed under this gauge
symmetry as $\psi_l(t)\rightarrow \exp{[i\alpha_l(0)]}\psi_l(t)$
independently of $t$. The product $\psi^{\dag}_l(0)\psi_l(t)$ is
thus manifestly gauge-invariant. Its phase after subtracting the
gauge-invariant ``dynamical phase" (DP) $\int_0^T
dtw_l^{\dagger}(t)\hat{h}w_l(t)$ becomes
\begin{eqnarray}\label{eq-GP}
\beta_l ={\rm arg}\left\{
w^{\dag}_l(0)w_l(T)\exp\left[i\int_{0}^{T}dt
w_l^{\dagger}(t)i\partial_{t}w_l(t)\right]\right\},
\end{eqnarray}
which is also manifestly gauge-invariant. This  $\beta_l$ is
understood as the holonomy of the {\em basis vector} associated with
exact hidden local symmetry (\ref{eq-hiddenlocal}) for all GPs,
either adiabatic or nonadiabatic, as explained in detail in
\cite{2007Fujikawa}. This construction is a generalization of BP for
the generic case $E_l\neq 0$, for which the Schr\"{o}dinger
amplitude does not satisfy the parallel transport condition
\cite{simon} but the basis vector can satisfy the parallel transport
condition with the help of gauge symmetry (\ref{eq-hiddenlocal})
\cite{2007Fujikawa}. For the noncyclic case, one can still identify
Eq.(\ref{eq-GP}) as a gauge-invariant noncyclic GP
\cite{1988Samuel}.

We here briefly compare the above unified formulation of GPs to the
conventional formulation where the adiabatic phase is defined to be
invariant under the symmetry identical to the above hidden symmetry
(\ref{eq-hiddenlocal}), whereas the nonadiabatic phase is defined to
be invariant in the so-called projective Hilbert space with the
equivalence class $\{e^{i\alpha(t)}\psi(t)\}$
\cite{1987Aharonov,1988Samuel}. As a consequence, the gauge
invariant nonadiabatic phase $\beta ={\rm
arg}\{\psi^{\dag}(0)\psi(T)\exp[i\int_{0}^{T}dt
\psi^{\dag}(t)i\partial_{t}\psi(t)]\}$ \cite{1987Aharonov,singh} is
nonlocal and nonlinear in the Schr\"{o}dinger amplitude $\psi(t)$,
which causes certain complications as was noted by Marzlin et al.
\cite{2004Marzlin}. In contrast, our $\beta_l$ in Eq. (\ref{eq-GP}),
which numerically agrees with the Aharonov-Anandan $\beta$ when one
uses exact solution (\ref{eq-exactamplitude}) in $\beta$, is
bilinear in the Schr\"{o}dinger amplitude.

\section{Geometric Phases in dissipation \label{sec-GP}}
Coming back to the qubit in a noisy environment, action
(\ref{eq-action1}) after taking the environment into consideration
is written in the second-quantized formulation as
\begin{eqnarray}\label{eq-action}
S&=&\int dt\Big\{\hat{\psi}^\dag(t)\Big(i\frac{\partial}{\partial
t}+{\bm B}\cdot {\bm
\sigma}/2\Big)\hat{\psi}(t)+\sum_\alpha\Big(\hat{p}_{\alpha}
\dot{\hat{x}}_{\alpha}\nonumber\\
&&-\frac{\hat{p}_\alpha^2}{2m_\alpha}
-\frac{m_\alpha\omega_\alpha^2}{2}\hat{x}_\alpha^2\Big)
-\hat{\psi}^\dag(t) \sigma_z\hat{\psi}(t)\sum_\alpha
C_\alpha\hat{x}_\alpha\Big\},\nonumber\\
\end{eqnarray}
where the environment is effectively described by an infinite number of bosonic oscillations \cite{footnote}. Note that we work on the case of the vanishing temperature $T=0$ in the present paper.
Accordingly, the effective Hamiltonian after considering dissipation
turns to be
\begin{eqnarray}\label{eff-hamiltonian}
\hat{H}_{\rm{eff}}&=&\sum_{l=\pm} E_l \hat{c}_l^\dag
\hat{c}_l+\sum_\alpha\omega_\alpha\left(\hat{a}_\alpha^\dag
\hat{a}_\alpha+\frac{1}{2}\right)\\
&&+\sum_{\alpha,l,m}C_\alpha \frac{i}{\sqrt{2m_\alpha\omega_\alpha}}
(\hat{a}_\alpha-\hat{a}_\alpha^\dag)(w_m^\dag\sigma_{z}
w_l)\hat{c}_m^\dag \hat{c}_l,\nonumber\label{eq-EffHam}
\end{eqnarray}
where
$\hat{x}_\alpha=\frac{i}{\sqrt{2m_\alpha\omega_\alpha}}(\hat{a}_\alpha
-\hat{a}_\alpha^\dagger)$,
$\hat{p}_\alpha=\sqrt{\frac{m_\alpha\omega_\alpha}{2}}(\hat{a}_\alpha
+\hat{a}_\alpha^\dagger)$, and the time-independent effective energy
eigenvalues $E_l$ have been given in Eq. (\ref{eq-Epm}). The exact
state vector for the qubit with the initial condition
$\psi_l(0)=w_l(0)$ is given by $\psi_{l}(t)=\langle
0|\hat{\psi}(t)\hat{c}^{\dagger}_{l}(0) |0\rangle$, i.e.,
\begin{eqnarray}\label{exact-evolution}
\psi_l(t)&=&\sum_m w_m(t)\langle
m|\mathcal{T}\exp[-i\int_0^t\hat{\mathcal{H}}_{\rm{eff}}(t')dt']|l\rangle\nonumber\\
&=&\sum_m w_m(t)\langle m|U(t)|l\rangle.
\end{eqnarray}
In our specific example in Eq. (\ref{eff-hamiltonian}),
$\hat{\mathcal{H}}_{\rm{eff}}(t)$ is time independent and simplifies
calculations. After integrating out the environmental freedom, Eq.
(\ref{exact-evolution}) assumes the form
\begin{eqnarray}\label{eq-Amplitude}
\psi_l(t)&\simeq&\sum_{m=\pm}\langle m|\exp\Bigg\{-i\int_0^t
\Bigg[\sum_{k=\pm}
E_k\hat{c}_k^\dag(0)\hat{c}_k(0)\nonumber\\
&&-\sum_{k,k'}\Sigma_{kk'}(t')\hat{c}_k^\dag(0)\hat{c}_{k'}(0)\Bigg]dt'\Bigg\}|l\rangle
w_m(t),\qquad
\end{eqnarray}
in which the second term in the square brackets is the lowest-order
``self-energy" correction, to be evaluated below due to the
interaction with environment.

Actually, our primary object of interest is evolution operator
(\ref{exact-evolution}) defined in terms of effective Hamiltonian
(\ref{eff-hamiltonian}). One can then integrate out the bosonic
freedom, which is Gaussian, exactly in the path-integral
representation of the evolution operator,
\begin{eqnarray}\label{pathintegral}
\int \prod_{\alpha}{\cal D}x_{\alpha}{\cal D}p_{\alpha}{\cal
D}\psi{\cal D}\bar{\psi}\exp\{iS\},
\end{eqnarray}
where the action $S$ is given by Eq. (\ref{eq-action}). This
procedure is somewhat analogous to the Gaussian integral of the time
component of the electromagnetic field $A_{0}$ in quantum
electrodynamics defined by the Coulomb gauge. One then obtains a
four-fermion coupling analogous to the Coulomb interaction. But in
the present case the Coulomb potential is replaced by the bosonic
free propagator defined by (see Eq. (3) of \cite{1992Fujikawa})
\begin{eqnarray*}
\lefteqn{\sum_{\alpha,\beta}\langle{\cal
T}C_{\alpha}x_{\alpha}(t)C_{\beta}x_{\beta}(t^{\prime})\rangle=
\frac{1}{\pi}\int_{0}^{\omega_{c}}d\omega^{\prime}J(\omega^{\prime})}\nonumber\\
&&\times\int_{-\infty}^{\infty}\frac{d\omega}{2\pi}
\left(\frac{i}{\omega-\omega^{\prime}+i\epsilon}-\frac{i}{\omega+\omega^{\prime}-i\epsilon}\right)e^{-i\omega(t-t^{\prime})},
\end{eqnarray*}
where we replaced the summation over $\alpha$ in the $x$ propagator by an
effective spectral density,
\begin{eqnarray}\label{spectral}
J(\omega)=\frac{\pi}{2}\sum_\alpha \frac{C_\alpha^2}{m_\alpha
\omega_\alpha}\delta(\omega-\omega_\alpha).
\end{eqnarray}
This spectral density typically has a power-law behavior at low
frequencies \cite{1987Leggett}. Of particular interest is the Ohmic
dissipation, corresponding to a spectrum $J(\omega)=\eta\omega$,
which is linear at low frequencies up to some high-frequency cutoff
$\omega_c$ ($\omega_c> B$). The dimensionless parameter $\eta$
reflects the strength of dissipation. Here we concentrate on weak
dissipation, $\eta\ll 1$, since only this regime is relevant for
quantum-state engineering \cite{2001Makhlin}. One may then perform
the fermionic path integral corresponding to the lowest-order
perturbative correction to the fermion self-energy depicted in
Fig.1. Alternatively, one can perform the same calculation by using
the Dyson formula $\mathcal{S}=\mathcal{T}\exp[-i\int H_I(t)dt]$
 in the interaction picture.

\begin{figure}
\includegraphics[width=4cm]{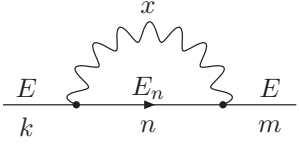}\\
\caption{The Feynman diagram for the self-energy correction $\Sigma_{mk}$ in the one-loop order: the solid line stands for the spin and the wavy line stands for the bosonic freedom.} \label{fig1}
\end{figure}

The actual evaluation of the Feynman diagram in Fig. \ref{fig1} for
the Ohmic case is straightforward by following the second-quantized
formulation of the Caldeira-Leggett model \cite{1992Fujikawa}. The
self-energy correction in the one-loop order is then given by (see
Eq. (5) of \cite{1992Fujikawa})
\begin{eqnarray}\label{eq-selfenergy}
\Sigma_{mk}^{(1)}&\equiv&\eta\sum_{l=\pm}(w_m^\dag{\sigma}_{z} w_l)(w_l^\dag\sigma_zw_k)\Big[i(E-E_l)\Theta(E-E_l)\nonumber\\
&&-\frac{(E_l-E)}{\pi}\ln\Big|\frac{\omega_c}{E-E_l}-1 \Big|\Big],
\end{eqnarray}
in which $\Theta$ is the step function. The first term in Eq.
(\ref{eq-selfenergy}), which is imaginary, gives the decay width of
the $k$th level as
\begin{equation}\label{width}
\frac{1}{2}\Gamma_k=\eta\sum_{l=\pm}\big|w_k^\dag\sigma_zw_l\big|^2(E_k-E_l)\Theta(E_k-E_l),
\end{equation}
which vanishes for the ground state due to the step function. It
indicates that the excited state decays to the ground state by
emitting soft bosonic excitations. Below we will see that it
characterizes the dephasing time scale of the spin system in
magnetic field. The second term of Eq. (\ref{eq-selfenergy}) is a
correction to the effective energy. As a result, the total effective
energy, to the order of $O(\eta)$, becomes
\begin{eqnarray}\label{1st-order-energy}
E_k^{\text{tot}}&=&E_k+\frac{\eta}{\pi}\sum_{l=\pm}\big|w_k^\dag\sigma_zw_l\big|^2(E_l-E_k)\nonumber\\
&&\times\ln\Big|\frac{\omega_c}{E_k-E_l}-1\Big|.
\end{eqnarray}

The Schr\"{o}dinger amplitude in Eq. (\ref{eq-Amplitude}) to the
accuracy of the lowest-order correction $O(\eta)$ is thus given by
using Eqs.(\ref{width}) and (\ref{1st-order-energy}) as
\begin{equation}\label{eq-psin}
\psi_l(t)\simeq  e^{-\Gamma_l
t/2}\exp[-i\int_0^tE_l^{\text{tot}}(t')dt'] w_l(t).
\end{equation}
It appears that the probability conservation for the higher-energy
state is violated in Eq. (\ref{eq-psin}). Mathematically, this
arises from the fact that we evaluated the ``persistent" amplitude
for the single-spin state under the influence of dissipation.
Evolution operator (\ref{eq-evolution}) is formally unitary in the
present case also. Thus the probability should be preserved in our
formulation and also in the formulation with the density matrix in
the Appendix.

In fact, the unitarity of the evolution operator,
$U^{\dagger}(t)U(t)=1$, is preserved if one evaluates
\begin{equation}\label{unitarity}
w_{l_{+}}(t)\langle l_{+}|U(t)|l_{+}\rangle
+w_{l_{-}}(t)\varphi_{B}(t)\langle l_{-};{\rm
bosons}|U(t)|l_{+}\rangle,
\end{equation}
since $w_{l_{-}}(t)\langle l_{-}|U(t)|l_{+}\rangle=0$ in the present
model, where $\langle l_{\pm}|$ respectively stand for the higher-
and lower-energy states of the spin, and $\langle l_{-};{\rm
bosons}|$ stand for the final states of the lower energy spin state
together with  soft bosonic excitations. (To be more precise, one
may write $\langle l_{\pm}|$ as $\langle l_{\pm}|\otimes\langle 0|$
and $\langle l_{-};{\rm bosons}|$ as $\langle l_{-}|\otimes \langle
{\rm bosons}|$ with $\langle 0|$ standing for the bosonic vacuum.
But we use the simplified notation in this paper.) Note that the
description in terms of effective spectral density (\ref{spectral})
is  different from the description in terms of a definite number of
bosonic quanta. In the spirit of the Caldeira-Leggett model
\cite{1981Caldeira}, we do not assign a physical significance to the
soft bosonic excitations \cite{footnote}. In fact, the actual cause
of the dissipation could be completely different from an ensemble of
an infinite number of harmonic oscillators, though we consider that
those harmonic oscillators, if suitably chosen, can mimic the actual
dissipation. The presence of the dissipation is thus manifested by
the decrease in the persistent amplitude $w_{l_{+}}(t)\langle
l_{+}|U(t)|l_{+}\rangle$. [In this respect, our bosonic freedom is
different from the time-component of the electromagnetic field in
QED, which does not influence the unitarity. In path integral
(\ref{pathintegral}), one needs to add a Schwinger's source term to
the bosonic freedom also to analyze the unitarity.]

Since the final states $w_{l_{-}}(t)\varphi_{B}(t)\langle l_{-};{\rm
bosons}|$ are orthogonal to both of  $w_{l_{+}}(t)\langle l_{+}|$
and $w_{l_{-}}(t)\langle l_{-}|$ in Eq. (\ref{unitarity}) and thus
do not interfere with these states when one considers only the
observables of spin freedom as in the present case, the decrease in
the persistent amplitude $w_{l_{+}}(t)\langle
l_{+}|U(t)|l_{+}\rangle$ is also understood as an indicator of the
quantum decoherence. See Eq. (\ref{eq-A3}) in the Appendix. The
inverse of the decay width in Eq. (\ref{width}) is thus related to
the two characteristic time scales of the qubits, namely,
$\tau_{\text{relax}}$ and $\tau_\varphi$ \cite{2008Clarke}. The
relaxation time scale $\tau_{\text{relax}}$ is the time required for
a qubit to relax from the first excited state to the ground state,
involving energy loss. The dephasing time scale $\tau_\varphi$ is
the time over which the off-diagonal [in the preferred basis vectors
$w_\pm(t)$] elements of the qubit's reduced density matrix decay to
zero in the formulation with the density matrix as in the Appendix.
In the present model, these two time scales are of the same order of
magnitude, and we choose $\tau_\varphi=2/\Gamma_{l{+}}$ by using the
decay width in Eq. (\ref{width}). We thus need to satisfy $T\ll
\tau_\varphi$ or
\begin{eqnarray}
\omega_{0}\gg \pi \Gamma_{l{+}},
\end{eqnarray}
to make a sensible measurement of  geometric phases, which are
defined for {\em pure} states. The gauge-invariant geometric phases
for mixed states can be defined, but their measurement is generally
rather indirect \cite{2007Fujikawa}.

\subsection{Adiabatic Berry phase\label{subsec-Adiabatic}}
\begin{figure}
  \includegraphics[width=8cm]{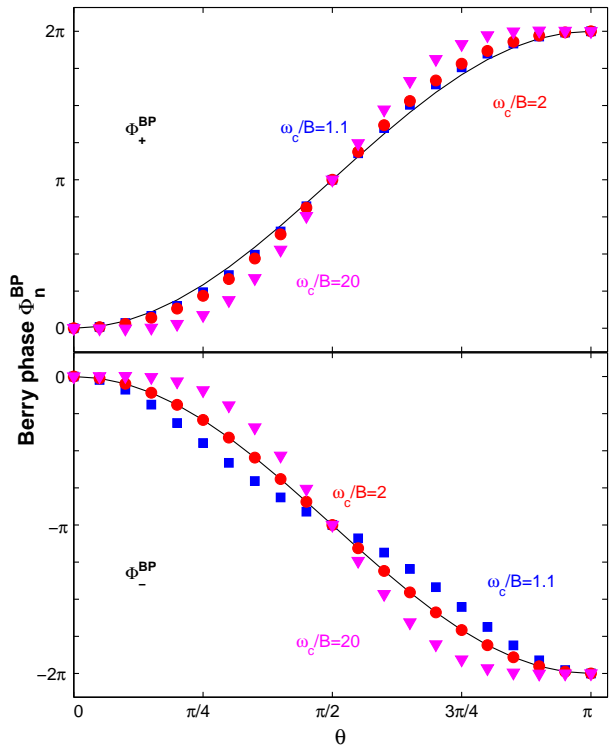}\\
\caption{(Color online) Berry phase $\Phi_n^{\text{BP}}$ for various
values of cutoff $\omega_c/B$. The black line corresponds to the
dissipationless case, while marker points represent the case with
dissipation. We set the dimensionless parameter $\eta=0.3$ and take
${\rm sign}(\omega_0)=+1$. The values of the period
$T=2\pi/\omega_{0}$ and the field strength $B/2\pi=50\times10^6$(Hz)
are taken from \cite{2007Leek}. }\label{fig2}
\end{figure}

We now come to the analysis of GPs with dissipation. For
definiteness, let us first analyze the environmental effects on the
adiabatic BP. In the adiabatic limit
$\theta_0\simeq\sin\theta\omega_0/B\rightarrow0$ in Eq.
(\ref{eq-Epm}), one has the non-vanishing quantities
\begin{eqnarray*}
\tau_\varphi^{-1}&=&\eta\sin^2\theta(B-\omega_0\cos\theta),\nonumber\\
E_{\pm}^{\text{tot}}&=&\mp
\frac{1}{2}B-\frac{1}{2}\omega_0(1\pm\cos\theta)\pm\frac{\eta}{\pi}\sin^2\theta
(B\nonumber\\
&&-\omega_0\cos\theta)\ln\Big|1\pm\frac{\omega_c}{B+\omega_0\cos\theta}\Big|.
\label{eq-concrete}
\end{eqnarray*}
Here we mention that to derive above $\tau_\varphi^{-1}$, we first
set $\theta_0\simeq\sin\theta\omega_0/B$ and then let
$\omega_0/B\rightarrow0$ in formula (\ref{width}). For
$|\omega_0|\ll B<\omega_c£¡$, the logarithm in the last term is
nearly independent of geometry and its sign is very important
because it determines whether the correction term is added or
subtracted. After evolving one cycle $T$ ($2\pi/B\ll T\ll
\tau_\varphi$), if one defines the phase
$\Phi_l=\Phi_l^{\text{DP}}+\Phi_l^{\text{BP}}=\int_0^TE_l^{\text{tot}}(t)dt$
up to $2\pi n$ ($n=$integer), we may identify  DP and BP
respectively as
\begin{eqnarray*}\label{eq-BP}
\Phi^{\text{DP}}_\pm&=&\mp \frac{1}{2}BT\pm\frac{\eta}{\pi}\sin^2\theta B T\ln\left|1\pm\frac{\omega_c}{B}\right|,\\
\Phi^{\text{BP}}_\pm&=&\mathrm{sgn}(\omega_0)\left[\pm\Omega\mp2\eta\sin^2\theta\cos\theta\ln\left|1\pm\frac{\omega_c}{B}\right|\right],\nonumber
\end{eqnarray*}
where the solid angle $\Omega=\pi(1-\cos\theta)$. The first terms in
both DP and BP above are the ones in the absence of noise, while the
second terms arise from the coupling between qubit and environment.
Our results on the correction of BP and the geometric dephasing
factor are consistent with those found by Whitney and co-workers
\cite{2003Whitney,2005Whitney}.

If $\omega_c<2B$, the BP correction of the ground state would have
the same sign as that of the excited state. Otherwise, the sign will
be opposite. For an illustration, we plot $\Phi_\pm^{\text{BP}}$
with respect to the polar angle $\theta$ for various values of the
cutoff $\omega_c/B$ in Fig. \ref{fig2}.

In passing, for a super-Ohmic case $J(\omega)=\eta \omega^3$ the
dephasing time scale is given by $\tau_\varphi^{-1}=\eta
\sin^2\theta B^2(B+\omega_0\cos\theta)$ and the correction to BP
becomes $\Delta\Phi^{\text{BP}}_\pm=\mp{\rm
sign}(\omega_0)2\eta\sin^2\theta\cos\theta
\left(\omega_c^2/2-B^2\ln|1\pm\omega_c/B|\right)$ in the adiabatic
limit.

\subsection{Nonadiabatic geometrical phase\label{subsec-nonadiabatic}}
\begin{figure}[tbph]
  \includegraphics[width=8cm]{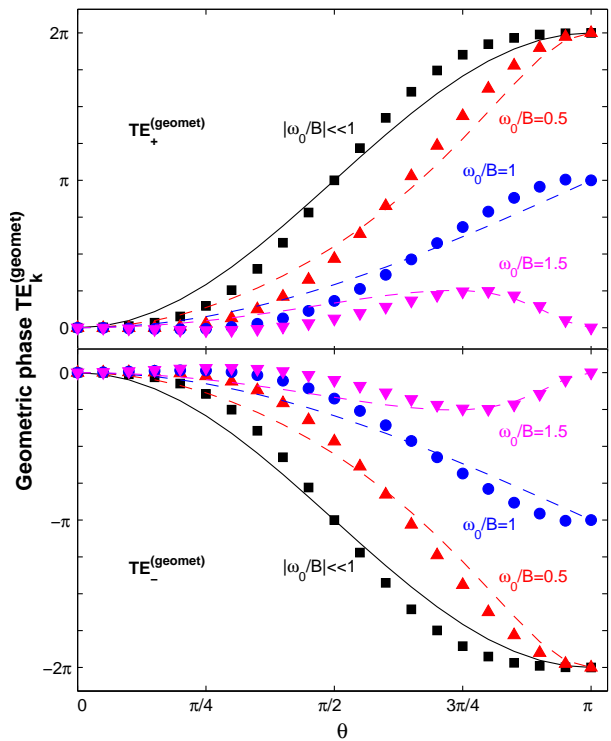}\\
\caption{(Color online) Non-adiabatic geometric phases for different
evolution rates $\omega_0/B$, in the presence or absence of
environment. All lines including solid and dashed ones correspond to
the nonadiabatic GPs without environment coupling (i.e.,
$\Omega_k$), while all marker points represent GPs,
$TE_k^{(\text{geomet})}$, with environment coupling. The adiabatic
limit of $|\omega_0/B|\ll1$ is shown by black color. We set the
dimensionless parameter $\eta=0.3$ and take ${\rm
sign}(\omega_0)=+1$. Other parameters are chosen as
$B/2\pi=50\times10^6$(Hz) and $\omega_c=3B$.}\label{fig3}
\end{figure}%

We emphasize that our formula (\ref{eq-psin}) is valid for both of
the adiabatic and nonadiabatic cases. Thus the modification of
nonadiabatic (Aharonov-Anandan) GP due to the dissipation is
analyzed quantitatively. A technical issue involved here is that a
clear separation of dissipation-induced GP from dissipation-induced
DP is not possible; One already recognizes this tendency even in the
case without dissipation in Eq. (\ref{eq-weigen}). We here
tentatively employ the following procedure for identifying the
geometric phase away from the ideal adiabaticity with dissipation.
As for the phase without dissipation, we adopt the separation in Eq.
(\ref{eq-weigen}). For the phase induced by dissipation, we define
the geometric part by
\begin{eqnarray*}
T\Delta E_{k}^{({\rm geomet})}=T [\Delta E_{k}(\omega_{0})- \Delta
E_{k}(\omega_{0}=0)],
\end{eqnarray*}
where $\Delta E_{k}(\omega_{0})$ stands for the second term in Eq.
(\ref{1st-order-energy}) proportional to $\eta$, and
$T=2\pi/|\omega_{0}|$. This identification may be reasonable if both
of $\omega_{0}$ and $\eta$ are small. Thus the total geometric phase
is
\begin{eqnarray*}
TE_{k}^{({\rm geomet})}\equiv\Omega_{k}+T\Delta E_{k}^{({\rm
geomet})},
\end{eqnarray*}
which  is illustrated in Fig. \ref{fig3} with respect to the
variable $\theta$. Here $\Omega_k=\pm\Omega_{\theta_{0}}$ stands for
the nonadiabatic GP without dissipation obtained by multiplying the
period $T$ to the second ``geometric energy" in Eq.
(\ref{eq-weigen}) and
$\Omega_{\theta_{0}}=\pi[1-\cos(\theta-\theta_0)]$ stands for the
(nonadiabatic) solid angle with $\theta_0$ defined in Eq.
(\ref{eq-xi}). When taking the adiabatic limit of
$|\omega_0/B|\ll1$, $TE_k^{(\text{geomet})}$ approaches the
adiabatic Berry phase with dissipation. When taking $\eta=0$,
$TE_k^{(\text{geomet})}$ is reduced to the nonadiabatic GP
$\Omega_k=\pm\Omega_{\theta_{0}}$ without dissipation.

Figure \ref{fig3} illustrates how the observed GPs, with and without
the effects of environment, deviate from the supposedly topological
BP, $\pm\pi(1-\cos\theta)$, when the variables $\omega_{0}$ and
$\eta$ deviate away from negligibly small values. It can be seen
that the nonadiabatic GP is affected sensitively by the change in
the dynamical parameter $\omega_0/B$, while the adiabatic BP is
supposed to be immune to small fluctuations of $\omega_0/B$. This is
related to the basic issue of the intrinsic robustness of BP. When
$\omega_0/B$ becomes big ({\em e.g.}, $>1$ in our choice of the
parameter values), the non-monotonous magenta dashed lines localized
near zero in Fig. \ref{fig3} show that the dissipationless
nonadiabatic GP $\Omega_k$ tends to be restricted in the
neighborhood of zero, the trivial value of GP.

In the transitional region from adiabatic to nonadiabatic limit, the
effect of environment on GP changes as $\theta$ increases and keeps
the same pattern for diverse $\omega_0/B$, as indicated by the red
and blue lines in Fig. \ref{fig3}. The red dashed lines,
corresponding to a relatively small value of $\omega_0/B$, tends to
recover the adiabatic case shown in  Fig. \ref{fig2}, while the blue
dashed lines away from both the adiabatic and nonadiabatic limits
reflect the transitional region. One may note that at $\theta=\pi$
GP takes one of values of $0$, $\pm \pi$, and $\pm2\pi$, which
sharply depends on the value of $\omega_0/B$. This can be easily
understood from the definition of $\theta_0$ in Eq. (\ref{eq-xi}).
In addition, it should be mentioned that for a fixed value of
$\omega_0/B$ the $TE_+^{(\text{geomet})}$ and
$TE_-^{(\text{geomet})}$ are not symmetric about the axis of
$TE_k^{(\text{geomet})}=0$ due to the existence of the logarithmic
term in Eq. (\ref{1st-order-energy}). As for the role of the
coupling strength $\eta$, it determines the magnitude of deviations,
while having little effect on the pattern of deviations at least in
the perturbative domain.

Finally, let us briefly analyze the geometric dephasing factor for
the nonadiabatic case. From Eq. (\ref{width}), the nonvanishing
decay width is given by
\begin{equation*}
\frac{1}{2}\Gamma_-=\eta\sin^2(\theta-\theta_0)[B\cos\theta_0+\omega_0\cos(\theta-\theta_0)].
\end{equation*}
Its geometric dependence is complicated because of  the involvement
of $\theta_0$ through Eq. (\ref{eq-xi}). At the adiabatic limit, it
returns to the $\tau_\varphi^{-1}$ discussed in Sec.
\ref{subsec-Adiabatic}. At the nonadiabatic limit, we have
$(\theta-\theta_0)\sim0$ and thus realize a dephasing-free
situation. Unfortunately, the limit corresponds to the trivial
geometric phase.

\section{Conclusion \label{sec-con}}
We have presented a gauge-invariant formulation of GPs for a
two-level system in dissipative environment, which is applicable to
both of adiabatic and nonadiabatic cases.  Our formulation may be
useful for understanding the experimental observation of BP such as
in a recent superconducting qubit \cite{2007Leek} and will provide a
starting point for the future quantitative analysis of the basic
issues of intrinsic robustness \cite{2003Blais} and  the behavior of
GP in the transitional region away from ideal adiabaticity. The
analysis of the region away from ideal adiabaticity is expected to
be crucial in any practical application of geometric phases.

\begin{acknowledgments}
We thank M. L. Ge for helpful discussions. This work was supported
by NSF of China (Grant No. 10575053) and LuiHui Center for Applied
Mathematics through the joint project of Nankai University and
Tianjin University.
\end{acknowledgments}

\appendix

\section{Density matrix}

One may be interested in the mixed states which are described by a density matrix in general.
 The time development of the density matrix $\rho$ is described  by the evolution operator as
\begin{eqnarray}
\rho(t)=U(t)\rho(0)U^{\dagger}(t),
\end{eqnarray}
 at the vanishing temperature $T=0$. The evaluation of $U(t)$ is thus sufficient.
For our purpose of simulating the dissipation by an infinite number
of harmonic oscillators, one may trace out
$\rho(t)=U(t)\rho(0)U^{\dagger}(t)$ with respect to all those final
states which contain bosonic excitations. In our simple model, the
bosonic excitations are included only in the lower-energy state of
the fermion as in Eq. (\ref{unitarity}). If one defines a state
\begin{eqnarray}
\psi(t)=a\psi_{l_{+}}(t)+b\psi_{l_{-}}(t),
\end{eqnarray}
with constants $a$ and $b$, which is no more cyclic even without
dissipation since $\psi(T)\neq \psi(0)$ up to a phase, and starts
with a pure state $\rho(0)=|\psi(0)\rangle\langle\psi(0)|$, one
obtains the mixed state after partial tracing of $\rho(t)$ over
those final bosonic excitations as
\begin{eqnarray} \label{eq-A3}
\tilde{\rho}(t)&=&|\psi(t)\rangle\langle\psi(t)|+|a|^{2}[|\psi_{l_{+}}(0)|^{2}-|\psi_{l_{+}}(t)|^{2}]\times\nonumber\\
&&|\psi_{l_{-}}(t)\rangle\langle\psi_{l_{-}}(t)|,
\end{eqnarray}
with $|\psi_{l_{+}}(0)|^{2}=1$. We here used the unitarity relation
in Eq. (\ref{unitarity}),
\begin{eqnarray}
&&|\psi_{l_{+}}(t)|^{2}+\sum_{\rm bosons}|\langle {l_{-}; \rm bosons}|U(t)|l_{+}\rangle|^{2}|\varphi_{B}(t)|^{2}
|w_{l_{-}}(t)|^{2}\nonumber\\
&&=|\psi_{l_{+}}(0)|^{2},
\end{eqnarray}
with $|w_{l_{-}}(t)|^{2}=1$. The first term in Eq. (\ref{eq-A3})
arises from the trivial bosonic vacuum and the second term in Eq.
(\ref{eq-A3}) from the nontrivial bosonic states. The density matrix
$\tilde{\rho}(t)$ approaches
\begin{equation}
\tilde{\rho}(t)\rightarrow
(|a|^{2}+|b|^{2})|\psi_{l_{-}}(t)\rangle\langle\psi_{l_{-}}(t)|,
\end{equation}
for  $t\rightarrow{\rm large}$. Thus the total trace $\mathrm{Tr}
\rho(t)=|a|^{2}+|b|^{2}$ is preserved during the time development in
the present normalization of the density matrix.

One may define gauge invariant geometric phases for the mixed state
described by this density matrix $\tilde{\rho}(t)$ following the
general formulation in \cite{2007Fujikawa}. An interesting case is
obtained if one sets $b=0$ and considers the time interval
$t_{0}\leq t\leq t_{0}+T$ with $t_{0}\neq 0$ for $\tilde{\rho}(t)$.


\begin{thebibliography}{30}
\bibitem{2008Clarke}
J. Clarke and F. K. Wilhelm, Nature (London) \textbf{453}, 1031 (2008).%
\bibitem{2007Majer}
J. Majer et al., Nature (London) \textbf{449}, 443 (2007).
\bibitem{2007sillanpaa}
 M. A. Sillanp\"{a}\"{a}, J. I. Park, and R. W. Simmonds, Nature (London) \textbf{449}, 438 (2007), and references therein.
\bibitem{2001Makhlin}
Y. Makhlin, G. Sch\"{o}n, and A. Shnirman, Rev. Mod. Phys.
\textbf{73}, 357 (2001).
\bibitem{2002Vion}
D. Vion et al., Science \textbf{296}, 886 (2002).
\bibitem{1984Berry}
M. V. Berry, Proc. R. Soc. London, Ser. A \textbf{392}, 45 (1984).
\bibitem{1992Anandan}
J. Anandan, Nature (London) \textbf{360}, 307 (1992).
\bibitem{2007Leek}
P. J. Leek, et al., Science \textbf{318}, 1889 (2007).
\bibitem{2008Mottonen}
M. M\"{o}tt\"{o}nen, J. J. Vartiainen, and J. P. Pekola, Phys. Rev.
Lett. \textbf{100}, 177201 (2008).
\bibitem{1999Nakamura}
Y. Nakamura, Yu. A. Pashkin, and J. S. Tsai, Nature (London)
\textbf{398}, 786 (1999).
\bibitem{1999Zanardi}
P. Zanardi and M. Rasetti, Phys. Lett. A \textbf{264}, 94 (1999).
\bibitem{2003Blais}
A. Blais and A. M. S. Tremblay, Phys. Rev. A \textbf{67}, 012308
(2003).

\bibitem{2003Chiara}
G. De Chiara and G. M. Palma, Phys. Rev. Lett. \textbf{91}, 090404
(2003).
\bibitem{2003Carollo}
A. Carollo, I. Fuentes-Guridi, M. F. Santos, and V. Vedral, Phys.
Rev. Lett. \textbf{90}, 160402 (2003).
\bibitem{2004Marzlin}
K.-P. Marzlin, S. Ghose, and B. C. Sanders, Phys. Rev. Lett.
\textbf{93}, 260402 (2004).
\bibitem{1981Caldeira}
A. O. Caldeira and A. J. Leggett, Phys. Rev. Lett. \textbf{46}, 211
(1981).
\bibitem{1987Leggett}
A. J. Leggett, et al., Rev. Mod. Phys. \textbf{59}, 1 (1987).
\bibitem{2003Whitney}
R. S. Whitney and Y. Gefen, Phys. Rev. Lett. \textbf{90}, 190402
(2003).
\bibitem{2005Whitney}
R. S. Whitney, Y. Makhlin, A. Shnirman, and Y. Gefen, Phys. Rev.
Lett. \textbf{94}, 070407 (2005).
\bibitem{1987Aharonov}
Y. Aharonov and J. Anandan, Phys. Rev. Lett. \textbf{58}, 1593
(1987).
\bibitem{1988Samuel}
J. Samuel and R. Bhandari, Phys. Rev. Lett. \textbf{60}, 2339
(1988).
\bibitem{2001Farhi}
E. Farhi et al., Science \textbf{292}, 472 (2001).
\bibitem{2006Nombardo}
F. C. Lombardo and P. I. Villar, Phys. Rev. A \textbf{74}, 042311
(2006).
\bibitem{2005Deguchi}
S. Deguchi and K. Fujikawa, Phys. Rev. A \textbf{72}, 012111 (2005).
\bibitem{2005Fujikawa}
K. Fujikawa, Phys. Rev. D \textbf{72}, 025009
(2005).
\bibitem{2007Fujikawa} K. Fujikawa, Ann. Phys. (N.Y.)
\textbf{322}, 1500 (2007).
\bibitem{2008Fujikawa}
K. Fujikawa, Phys. Rev. D \textbf{77}, 045006 (2008).
\bibitem{simon}
B. Simon, Phys. Rev. Lett. {\bf 51}, 2167 (1983).
\bibitem{singh}
K. Singh, D. M. Tong, K. Basu, J. L. Chen, and  J. F. Du, Phys. Rev.
A {\bf 67}, 032106 (2003).
\bibitem{footnote}
In the spirit of the Caldeira-Leggett model \cite{1981Caldeira}, we
treat the bosonic freedom as an auxiliary device to simulate the
dissipation. We do not take the infinite number of bosonic
excitations as  physical objects. See  K. Fujikawa, Phys. Rev. E
\textbf{57}, 5023 (1998).
\bibitem{1992Fujikawa}
K. Fujikawa, S. Iso, M. Sasaki, and H. Suzuki, Phys. Rev. Lett.
\textbf{68}, 1093 (1992).

\end{thebibliography}
\end{document}